\documentclass{edm_template}
\usepackage{float}
\begin{document}

\title{A Social Network Analysis on Blended Courses}

\numberofauthors{1}
\author{
\alignauthor
Niki Gitinabard, Linting Xue, Collin F. Lynch, Sarah Heckman, Tiffany Barnes\\
      \affaddr{North Carolina State University}\\
      \affaddr{Computer Science Department}\\
      \affaddr{Raleigh, NC, US}\\
      \email{\{ngitina, lxue3, cflycnh, sarah\_heckman, tmbarnes\}@ncsu.edu}
}

\maketitle
\begin{abstract}
The large-scale online management systems (e.g. Moodle), online web forums (e.g. Piazza), and online homework systems (e.g. WebAssign) have been widely used in the blended courses recently.  Instructors can use these systems to deliver class content and materials.  Students can communicate with the classmates, share the course materials, and discuss the course questions via the online forums.  With the increased use of the online systems, a large amount of students' interaction data has been collected. This data can be used to analyze students' learning behaviors and predict students' learning outcomes.  In this work, we collected students' interaction data in three different blended courses. We represented the data as directed graphs and investigated the correlation between the social graph properties and students' final grades. Our results showed that in all these classes,  students who asked more answers and received more feedbacks on the forum tend to obtain higher grades. The significance of this work is that we can use the results to encourage students to participate more in forums to learn the class materials better; we can also build a predictive model based on the social metrics to show us low performing students early in the semester. 
\end{abstract}


\section{Introduction}
Learning Management Systems (LMS) such as Moodle, online web forums such as Piazza and online homework systems such as WebAssign have become increasingly popular in recent years.  At the college level it is increasingly common for courses to be taught as blended courses with instructors using online tools to deliver materials, students completing assignments online, and participants communicating via online forums.  It is often easier for students and instructors to interact via the online systems because they don't have to be at the same place as the instructor to access course material and they can ask questions or submit assignments, often with automatic feedback, at any time.  These systems also collect more and more detailed information about students' study behaviors than that collected from traditional courses. In traditional face-to-face discussions it is difficult to track how students communicate or whom they communicate with.  This is also true for cases where they use direct messaging such as email or text messages.  This is why prior work on social relationships in courses has relied on interviews and self-report data (e.g. \cite{Fire}). In blended courses, however, we  can often  access students'  forum discussions as well as their homework submissions and even file accesses all of which allow us to identify both explicit and implicit social interactions.  This in turn provides us with the opportunity to study students' social relationships and work habits on a more detailed level than before.

Prior research on social relationships in MOOCs has shown that metrics of social activity such as degree, centrality, etc. are correlated with both dropout and with student performance \cite{jiang14, hmelo14, brown15}. In MOOCs, all or almost all student interaction occurs online.  Students in the course may be many hundreds or thousands of miles apart.  They may operate on widely different schedules.  And they may have no other means of contact apart from the online forums.  While some students do take courses in tandem or arrange meetups such external interaction is comparatively rare relative to the class size. Thus prior researchers have assumed that all relevant interactions are found online.  In blended courses, however, this is not the case.  Blended courses still include a significant face-to-face component such as working lectures or office hours.  And students are enrolled at the same institution.  They typically operate on similar schedules, and they may have preexisting social relationships that predate the course.  Thus it is far easier for them to meet face to face or to communicate via other channels.   Thus it is not clear that the relatively strong results that have been found in MOOCs will hold for blended courses.

Our primary goal in this work is to determine whether or not existing social network metrics (i.e. in-degree, out-degree, betweenness centrality, hub-score, and authority-score), are correlated with student performance in blended courses.  In order to assess this, we collected students' interaction data from three different blended courses, two of which include a structured peer tutoring component.  Having collected this data we first studied the active and inactive users' performance and determined whether or not the differences between them were significant.  Then we encoded the available interaction data as a layered social network and studied whether or not the social metrics were correlated with  student grades.  Because our datasets include both different instructors and different levels of interaction we were able to study how these differences are reflected in students' observed social behavior. 

\section{Literature Review}
\subsection{MOOCs}
While early MOOCs provided by Coursera and others tended to follow a predictable pattern new structures have begun to emerge such as CMOOCs \cite{palermos}.  Moreover, MOOCs vary widely both in topic areas and in the extent to which they provide either certification or clear records of performance, and students vary in their reasons for enrolling in them as well as their expectations for course outcomes \cite{wang14,wang15}.  As a consequence, research on the relationship between students' online social behavior and course outcomes in MOOCs evidences conflicting results.  Ramesh et al., for example, found that students who spent more time on average in forums were more likely to earn certificates of completion \cite{ramesh13}.  Deboer et al. found a similar result for students who responded to more threads than their peers \cite{deboer13}, while Yang et al. found that students who started threads less frequently than their peers were less likely to do so \cite{yang15}.  Yang et. al. also showed that the students who received more replies on their posts were more likely to complete the courses.  Andres et. al., by contrast, attempted to replicate some of the results found by De-Boer et al., Ramesh et al., and Yang et al. on a different dataset, and found that not all of the previous results held \cite{andres}.

Prior researchers have also studied the correlation between students' social positioning and their course grades.  Jiang et. al. analysed 2 different MOOCs and found that in one of the classes (``Intermediate Algebra'') there was a significant correlation between the students' grades and both their degree centrality within the network as well as their betweenness centrality.  That result did not hold however, when they examined the second class in their dataset (``Fundamentals of Personal Financial Planning''). They suggested that this might be due to the difference in course topics \cite{jiang14}. Hmelo-Silver et. al. showed that students with high authority and hub scores and a high number of formed relationships in the graph predict lower attrition \cite{hmelo14}. Yang et. al. created a survival model based on features reflecting students' behaviors \cite{yang13} to predict student dropout rates. These features were related to posting behavior (e.g. post length), joining time (e.g. the week in which they joined the MOOC), and social network behavior (e.g. authority score). They found that among all social network features, cohort, post duration, and authority strongly affected dropout along the way during participation in MOOCs. Kovanovic et. al. categorized students' ``social presence'' as \emph{affective} (e.g. expression of emotion), \emph{interactive} (e.g. continuing a thread), and \emph{cohesive} (e.g. vocatives)  by manually labeling the forum posts. They found that the betweenness centrality, in-degree, and out-degree of each student in the social graph generated by their replies on the discussion forum were significantly predicted from the presence metrics. In other words, the value of interactive social presence was strongly associated with the social graph metrics \cite{kovanovic14}.  

\subsection{Blended Courses}
Online discussion forums are a popular component of blended courses as they provide an easy mechanism for persistent communication shared question answering for students and instructors. Instructors can post materials for general use online and can answer shared questions once rather than repeating them for each student in office hours.  Students in turn can pose questions and receive feedback and explanations from classmates, TAs or instructors at any time of day. 

Research on blended courses, by contrast, has been limited. Vellukunnel et. al. studied a blended course on Java Programming Concepts, also a dataset in our work.  They showed that the students typically used the discussion forum for logistics and for relatively shallow questions \cite{vellukunnel17}. They also observed that the average grade of students who asked at least one question on the forum was significantly higher than those who did not. 

Thus the results of prior work are complex and are often contradictory.  One potential explanation for this variation lies in course content, another lies in the variable motivations of students, and still a third in the instructor's own instructional methodology.  While some results do generalize across MOOCs or blended courses it is clear that further research is required to determine what if any of the prior results generalize to our present context.  Further research is also required to assess the impact of instructional strategy and content on the observed social behaviors.

\section{Dataset Information}

In this paper we report on studies of three distinct courses, ``Discrete Math-2013'', ``Discrete Math-2015'' and ``Java Programming Concepts-2015''. All three are undergraduate computer science courses, offered at NC State and include significant blended components. Piazza is the system used in all these courses as a discussion forum. It lets students start a new thread for each question they want to ask, or information they like to share. When someone goes on a thread, they will be able to see in the same page the question, all the replies other users have given to that, and all the feedback given on each reply. In all these courses, participation on Piazza was not mandatory but the students were highly encouraged not to use email for asking questions and to post their questions on Piazza discussion forum. Discrete Math-2015 and Java Programming Concepts-2015 occurred contemporaneously during the Fall 2015 semester while Discrete Math-2013, a previous offering of Discrete Math-2015, was offered in Fall 2013. Discrete math covers topics such as propositional logic, predicate calculus, methods of proof, elementary set theory, and  the analysis of algorithms and asymptotic growth of functions. The emphasis in Java programming concepts is placed on software system design and testing; encapsulation; polymorphism; composition; inheritance; linear data structures; specification and implementation of finite-state machines; interpretation of inductive definitions (functions and data types); and resource management. General statistics for these courses is shown in Table~\ref{tab:stats}. Our datasets include the Piazza discussions and final grades for all the classes.

\begin{table}
\centering

\caption{Statistics of Each Class}
\label{tab:stats}
\begin{tabular}{|l|c|c|c|} \hline
\textbf{Class} & \textbf{DM 2013} & \textbf{DM 2015} & \textbf{Java}\\ \hline
Total Students & 251 & 255 & 181 \\ 
Average Grade & 81.2 & 87.6 &  79.7 \\
Participation Rate & 64.1\% & 65.8\% & 79.0\%\\
Average Piazza Actions & 11.79 & 9.44 & 14.68\\

\hline
\end{tabular}
\end{table}

\subsection{Discrete Math Classes}

Discrete Math-2013 had a total of 251, and Discrete Math-2015 had a total of 255 students. Each semester consisted of two sections taught by two different instructors and 5 shared teaching assistants. The average final grade in Discrete Math-2013 was 81.2 ans it was 87.6 in the 2015 session. Both sections used the same Moodle webpage for sharing assignments, a Piazza forum for discussions, and both used WebAssign and hand-graded homeworks.  After the first three assignments, students with 90\% or better average on those were given a chance to act as a peer tutor. If they completed ten hours of scheduled support including face to face office hours and online question answering, they would be exempt from taking the final exam. Note that the instructor indicated that she chose to answer less questions in the Discrete Math-2015 class, trying to give the peer tutors more chances to answer their classmates. 

\subsection{Java Programming Concepts Class}

This class had a total of 181 students in two different sections with different instructors but the same teaching assistants. There was also a distance education section for this course.  We omitted the distance education students from all the steps of our analysis as we only focus on students who can engage in face-to-face interactions. This class used Piazza for discussions, Moodle for course materials, Github for group projects, and Jenkins for automated code evaluation.

\section{Methods}

\subsection{Defining the Social Graph}
We constructed a directed social graph based on the online forum discussion data for each class. Each node in the graph represents a participant in the course(\emph{Instructor},\emph{TA}, or \emph{Student}).  In the discrete math class some students were also classified as\emph{Peer tutor}. In the discrete math 2013 class, posting completely anonymously was permitted, this produced unknown users which was removed from the further analysis. The arcs between users are defined using a similar method as in Brown et. al. \cite{brown15}. A directed edge (u, v) was added between users $u$ and $v$ for each instance where $u$ replied to a thread following $v$.  As in that case, e assume any user who posts on a thread has read all the previous replies and is replying back to all of them. We include all of the users in this graph so that it includes all the interactions of students, with other peers or with the teaching staff. We then aggregated the links between each two students and kept only one weighted edge between them in each direction.

\subsection{Comparison between Active and Non-active Student Grades}
In order to determine if participation in forums is effective on student outcomes, first we grouped students of each class into two groups of ``Active'' and ``Non-active''. Active students are students who contributed to the online forum. Non-active students are the ones who did not make any contribution. Note that we do not have access to the view data, so the non-active students may have viewed a large number of posts without being recorded. We then calculated the average grade of each group, and also conducted a T-Test between the group grades to see if the difference is significant.

\subsection{Social Metrics}
In order to test the correlation between social metrics and students' learning outcomes, we removed the non-active students from the graphs. We then calculated different graph and social metrics on the resulting graphs. These metrics consist of in-degree, out-degree, betweenness centrality, hub score, and authority score. Betweenness centrality is defined as a measure of the extent to which a vertex lies on the paths between others \cite{freeman}. Hubs and authorities are defined as a mutually reinforcing relationship: a good hub is a node that points to many good authorities; a good authority is a node that is pointed to by many good hubs \cite{kleinberg}. The in degree shows the number of replies and feedbacks the student has received; out degree indicates the number of replies and feedbacks the student has given; betweenness centrality tells how important this user is in connecting different users to each other, nodes with high betweenness are described as having some degree of control over the communication of others \cite{freeman}; users with high hub scores are those who frequently respond to the other active learners that post questions on the forum; students with high authority scores are the ones that receive most replies from the hub students. We will rename the hub score as ``help providing score'' and the authority score as ``help receiving score'' in our context to avoid confusion. In the end, we calculated the correlation between each of these metrics and student grades to see if different types of activity are correlated with student performance.

\section{Results and Discussion}
\begin{table}
\centering

\caption{ Average Grades for Active and Non-active Groups and T-Test p-values between the Two Groups' Grades}
\label{tab:ttest}
\begin{tabular}{|l|c|c|c|} \hline
Class & active & non-active  & p-value \\ \hline
Discrete Math 2013 & 89.25 & 58.37 & \textbf{1.09e-13} \\
Discrete Math 2015 & 90.74 & 81.61 & \textbf{1.43e-06}\\
Java Programming & 85.35 & 58.37  & \textbf{4.07e-09}\\ \hline
\end{tabular}
\end{table}

\begin{table}
\centering

\caption{Spearman Correlation between Grades and Graph Metrics}
\label{tab:correlations}
\begin{tabular}{|l|c|c|}
\multicolumn{1}{c}{} \\
\multicolumn{3}{c}{\textbf{Discrete Math 2013}} \\
 \hline
Class & correlation & p-value\\ \hline
In Degree & 0.32 & \textbf{2.91e-05}\\
Out Degree & 0.35 & \textbf{6.57e-06}\\
Betweenness Centrality & 0.33 & \textbf{1.52e-05}\\ 
Help Providing Score & 0.35 & \textbf{4.80e-06}\\
Help Receiving Score & 0.32 &\textbf{ 2.93e-05}\\\hline
\multicolumn{1}{c}{} \\
 \multicolumn{3}{c}{\textbf{Discrete Math 2015}} \\ 
\hline
 Class & correlation & p-value\\ \hline
In Degree & 0.22 & \textbf{0.0070}\\
Out Degree & 0.23 & \textbf{0.0061}\\
Betweenness Centrality & 0.22 & \textbf{0.0075}\\

Help Providing Score & 0.24 & \textbf{0.0033}\\
Help Receiving Score & 0.26 & \textbf{0.0015}\\\hline

\multicolumn{1}{c}{} \\
 \multicolumn{3}{c}{\textbf{Java Programming}} \\ 
\hline
 Class & correlation & p-value\\ \hline
In Degree & 0.26 & \textbf{0.005}\\
Out Degree & 0.21 & \textbf{0.022}\\
Betweenness Centrality & 0.28 & \textbf{0.002}\\ 
Help Providing Score & 0.09 & 0.34\\
Help Receiving Score & 0.07 & 0.46\\
 \hline
\end{tabular}
\end{table}

Table ~\ref{tab:ttest} shows the results of a T-Test comparing the active and non-active student grades and the average grade for each group in three classes.  As the table indicates, the average grades for the students who participated in the discussion forum are significantly higher than the average grades for the students who did not. One possible reason is that the students in the active group engaged more on the forum by replying other students' questions or receiving feedback from other active learners, which eventually helped them to learn and to obtain better grades. However, the non-active students who participated less on the forum may not understand the class materials very well, which results in getting lower grades. This result not does show causality, but it seems that either lower performing students participate less on the forum, or participation on the forum has helped those students to do better.

The results of the correlations between the five social metrics and students grades are shown in Table~\ref{tab:correlations}, in which bolded p-values (p<=0.05) in the right column means the correlations are significant. The results show that in all three classes, in degree, out degree, and betweenness centrality are positively and significantly correlated with the final grades. This indicates that the students, who received and posted more responses and who were more central in the graph (on the shortest path between more peers) performed better in the course. On the other hand, help providing score and help receiving score were only correlated with grades in the Discrete Math classes. This shows that the students who were connected to many popular users (the ones posting more questions or replies) are likely to obtain better grades in the Discrete Math course but not in the Java-Concepts class. Also, the correlations on the Discrete Math 2015 class are weaker than the 2013 class. This makes more investigation necessary, to find out the difference between these classes that might cause this result.

\begin{figure}[H]
  \centering

  \includegraphics[width=0.5\textwidth]{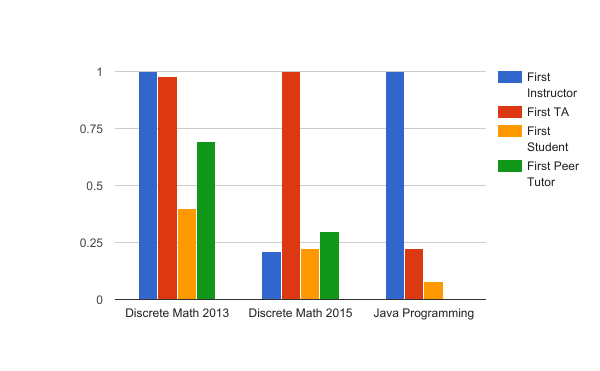}
  \caption{Top help-providing scores for each role in each class}
    \label{hub}
\end{figure}

The main difference between the two classes is the existence of peer tutors.  As noted above, in the discrete math class, higher performing students were required to complete ten hours of peer tutoring in forms of office hours or answering forum questions to be exempt from taking the final exam. The highest help providing scores among different roles in each class is shown in Figure~\ref{hub}. If we take a closer look at the biggest help providers in these classes, in all the classes the first one is an instructor or a TA. This finding is not surprising and shows that the teaching staff answered most of the questions of the most active users. The interesting part is the way these numbers decrease after that. In the discrete math 2013 class, the help providing scores seem to decrease more gradually than in the Java Programming and the Discrete Math 2015 class. The top help providing student in the Java class has a score of 0.078, which is even lower than 10\% of the top help providing user, who is the instructor in this case. In that class, all the students have very small help providing scores, which shows that the students were not active in answering their peers' questions. In this case, even the top help providing students do not have a much different score than the lowest help providing students. As a consequence of this gap and when considering the definition of help-receiving score, the users with the highest help-receiving score are mainly the ones most connected to the instructor, which leads to the poor correlation results. In order to better understand these results, we interviewed the course instructor.  She did not find it surprising. She noted that most of the questions on the forum were answered by herself or by the other teaching staff.

Using peer tutors and encouraging them to answer more questions seems to make this condition a different from the discrete math classes. In the 2013 class, the most \textit{help providing} student has a 0.69 score which is close to 70\% of the \textit{top help provider}, who is the instructor in this case. The highest help providing score among students of the Discrete Math 2015 class is not as high, but it is still close to 30\% of the top help provider. While this study does not show causality, we observe that in the classes with peer tutors, students are more engaged in answering their peer classmates' questions. In these classes, being connected to more active users is significantly correlated with student grades.

The difference between the two Discrete Math classes, as noted by the instructor, was that in the second one (2015), the instructor intentionally delayed her responses to the students, so that the students would feel encouraged to engage in the discussion. However, the comparison between the students' participation in these classes in Figure~\ref{hub} shows that this was not always the case. As we observed, in the second class one TA effectively took the place of the instructor in answering most of the questions. That decreased student participation in answering each others' questions. More research is needed to determine why replacing the instructor with a TA decreased students' contributions. One possibility is that the quality of the answers given by these two users was quite different. The instructor might have posted more challenging replies that would still keep the students engaged in the conversation, rather than an exact answer to the question that closes the discussion.

\section{Conclusions}
In this work, we investigate how the students' social metrics are correlated learning outcomes in three blended courses. To do this, we represented students' online interaction data as directed graph.  Our research showed that in all of the blended courses of our study, the in degree, out degree and betweenness scores are correlated with students' final grades. That shows, students who get more answers from other users; who give more answers to the other users; and who are the central users connecting others together, tend to perform better than other students in the class. We also observed that in classes that used peer tutoring, students are more engaged in answering others' questions and in these classes, being connected to more active users is significantly correlated with final grades.

\section{Future Work}
Our future work will be focused on using students' online social behaviors early in the semester to build a model that predicts their final grades. That model can help instructors to find potentially lower performing students early in the semester and help them.  We will also work on checking the effectiveness of peer tutors in a more controlled manner. The test will be performed on a control and test group among the same course with the same instructors and teaching staff, where the only difference among them is the use of peer tutors. In that case we can find out if using peer tutors can boost up students' activities on forum.

These results can be checked for consistency in MOOCs (Massive Online Open Courses) as well. Since the nature of those courses are different and the students and instructors often do not have a chance to meet in person, the results might be different. Also, performing this study on more classes can give us a more general conclusion.

\section{Acknowledgments}
This work was supported by NSF grant \#1418269: ``Modeling Social
Interaction \& Performance in STEM Learning'' Yoav Bergner, Ryan
Baker, Danielle S. McNamara, \& Tiffany Barnes Co-PIs.

\bibliographystyle{abbrv}
\bibliography{references}

\end{document}